\documentclass[prd,letterpaper,preprint,showpacs,superscriptaddress,floatfix]{revtex4}

\usepackage{graphicx,psfrag,amsmath,amssymb,amsfonts,bbm,latexsym,color,dcolumn,bm}

\begin{document}

\title{On weak interactions as short-distance manifestations of gravity}

\author{Roberto Onofrio}
\email{onofrior@gmail.com}
 
\affiliation{Dipartimento di Fisica e Astronomia 'Galileo Galilei', 
Universit\`a di Padova, Via Marzolo 8, Padova 35131, Italy}

\affiliation{ITAMP, Harvard-Smithsonian Center for Astrophysics, 
60 Garden Street, Cambridge, MA 02138, USA}

\date{\today}

\begin{abstract}
We conjecture that weak interactions are peculiar manifestations of quantum gravity at the 
Fermi scale, and that the Fermi constant is related to the Newtonian constant of gravitation.
In this framework one may understand the violations of fundamental symmetries by the weak interactions, 
in particular parity violations, as due to fluctuations of the spacetime geometry at a Planck 
scale coinciding with the Fermi scale. As a consequence, gravitational phenomena should play 
a more important role in the microworld, and experimental settings are suggested to test this 
hypothesis. 
\end{abstract}


\maketitle

Progress in Physics is sometime stimulated by critically revising already collected facts 
under a novel perspective, and the emergence of previously unnoticed connections. 
As a chief example, the realization that static measurements of the vacuum permittivity 
$\epsilon_0$ and permeability $\mu_0$ constants are related to the speed of light $c$ 
- determined in a setting perceived as completely different - as $c={(\mu_0 \epsilon_0)}^{-1/2}$, 
stimulated the development of a unified electromagnetic theory. 

In this spirit, in this communication we discuss a somewhat similar situation with an attempt 
to relate weak and gravitational interactions, which naturally share various features. 
First, they are universal, acting on all fundamental particles discovered so far, with the 
only exception of the gluon which does not seem to interact weakly, although it would be 
extremely difficult to experimentally evidence possible $gW^{\pm}$ and $ggZ^0$ vertices. 
Second, they have peculiar behavior with respect to external symmetries: the gravitational 
interaction, in the Einsteinian view, is the manifestation of curved spacetime itself, and  
weak interactions manifest parity violations, {\it i.e.} their structure is also quite sensitive 
to the spacetime. Third, they have {\sl small} interaction couplings, to be meant in a qualitative 
way that they can both be neglected when dealing with extremely short timescales. Finally, their 
coupling constants, the Newton gravitational constant $G_N$ and the Fermi constant $G_F$, have 
physical dimensions. These features are quite distinctive from the ones characteristic of electromagnetic 
and color interactions, which are instead selectively discriminating among particles based on the 
presence of electric and color charges, do preserve all external symmetries, and have dimensionless 
coupling constants large enough to determine the short-time dynamics of nearly all their bound states.

Therefore, a sensible question to ask is whether a joint analysis of weak and gravitational 
interactions is possible. Previous attempts have focused on embedding weak interactions 
in a broader structure of spacetime, such as the Einstein-Cartan theory in which contact 
interactions mediated by spin and torsion emerge naturally \cite{Hehl,Hehl1,Desabbata,Desabbata1}, 
or in introducing {\sl graviweak} interactions allowing the gravitational interaction to break 
external symmetries \cite{Loskutov}. Such a geometrical approach to weak interactions has been recently 
reconsidered in the framework of the so-called $f(R)$ gravity \cite{Capozziello,Capozziello1}, since 
the latter allows for a running coupling constant potentially capable to match the coupling constant 
present in the contact interaction of the Einstein-Cartan theory with the phenomenologically 
determined Fermi constant of charged weak interactions.

More specifically, we aim to provide further arguments in favor of a {\sl geometrization} 
program of the weak interactions based upon simple dimensional and numerical arguments. 
The Fermi constant is 
\begin{equation}
G_F=1.166364 \times 10^{-5} \mathrm{GeV^{-2}} (\hbar c)^3=1.43583 \times 10^{-62}~ \mathrm{m^5~ kg~ s^{-2}},
\end{equation}
where the last expression has been obtained by using MKSA units, and the dimensions of $G_F$ 
are $[G_F]=L^5 M T^{-2}$. The Newton universal constant of gravitation is instead
\begin{equation}
G_N=6.67384 \times 10^{-11}~ \mathrm{m^3~ kg^{-1}~ s^{-2}},
\end{equation}
with dimensions $[G_N]=L^3 M^{-1} T^{-2}$. The ratio between the two constants has 
therefore dimensions $[G_F/G_N]=L^2 M^2$, which is also the dimension of the square 
of the ratio of the two fundamental constants $\hbar$ and $c$. From the dimensional 
viewpoint, we can therefore write a relationship $G_F= \xi (\hbar/c)^2 G_N$, where $\xi$ 
is a dimensionless constant to be determined. By substituting the experimentally determined 
values of $G_F, G_N, \hbar,$ and $c$ we obtain $\xi=1.73867 \times 10^{33}$.
We notice that $G_N$ is measured in a range larger than the cm scale, while $G_F$ is measured 
through decay or scattering processes occurring in the subnuclear world, {\it i.e.} at distances 
smaller than $10^{-15}$ m. It is therefore highly tempting to conjecture that the prefactor $\xi$ 
would actually become of order unity if $G_N$ were also measured at subnuclear distances. 
In other words, by reabsorbing $\xi$ into the Newton's constant, we may write instead 
(the presence of the $\sqrt{2}$ factor will become clear in Eq. (4)): 
\begin{equation}
G_F=\frac{1}{\sqrt{2}}\left(\frac{\hbar}{c}\right)^2 \tilde{G}_N,
\end{equation}
provided that $\tilde{G}_N=2.45885 \times 10^{33} G_N=1.641 \times 10^{23}~ \mathrm{m^3~ kg^{-1}~ s^{-2}}$ 
is a {\sl renormalized} Newton's constant as measurable at subnuclear distances. 
Equation (3) suggests an interpretation of weak interactions as corresponding to a specific class of 
short-distance gravitational phenomena occurring due to quantum relativistic fluctuations of spacetime, 
and disappearing in the classical, non relativistic limit ($\hbar \rightarrow 0$ and $c^{-1} 
\rightarrow 0$). 
While on macroscopic lengthscales the effect of metric fluctuations is negligible, at small distances 
the foamy structure of spacetime \cite{Wheeler,Hawking,Garay,Carlip} allows for large fluctuations of 
the geometry. This could possibly include also states with opposite parity and phenomena such as a 
{\sl local} non-conservation of the electric charge, which then allows charge transport between two 
locations of spacetime, corresponding to the propagation of intermediate charged vector bosons 
$W^{\pm}$ in the standard model. 

This interpretation is also consistent with the fact that the Planck length, usually defined as 
$\Lambda_\mathrm{P}=\sqrt{\hbar G_N/c^3} = 1.6162 \times 10^{-35}~ \mathrm{m}$, is accordingly 
boosted by a factor $(\sqrt{2}\xi)^{1/2}=4.95868 \times 10^{16}$ when replacing $G_N$ with 
$\tilde{G}_N$, leading to $\tilde{\Lambda}_\mathrm{P}=8.0142 \times 10^{-19}$~ m. 
The corresponding Planck energy is 
$\tilde{E}_\mathrm{P}=\hbar c/\tilde{\Lambda}_\mathrm{P}=3.9449 \times 10^{-8} \mathrm{J}= 246.221$ GeV.  
Therefore the Planck energy coincides with the vacuum expectation value $v$ of the Higgs field in the 
electroweak model, as seen by inverting Eq. (3), thereby avoiding the hierarchy problem 

\begin{equation}
\tilde{E}_\mathrm{P}=\left(\frac{\hbar c^5}{\tilde{G}_N}\right)^{1/2}=
\left(\frac{(\hbar c)^3}{\sqrt{2}G_F}\right)^{1/2}=v.
\end{equation}  

If all above has anything to do with reality, the Newtonian constant of gravitation should 
be running, reaching at the attometer scale a value about $2.46 \times 10^{33}$ times larger 
than its value measured at the macroscopic level, {\it i.e.} 33 orders of magnitude over 
13 orders of magnitude in lengthscale. We have no precedent of such a steep distance-dependence 
for other coupling constants, however our current phenomenological knowledge of gravity cannot 
rule out this possibility without further experimental scrutiny, since the upper bounds on 
gravitational or gravitational-like forces are rather loose \cite{Antoniadis}. 
It is worth to point out that, with different motivations, the concept of {\sl strong} gravity 
has appeared from time to time in the literature, especially in connection with the possibility 
that gravity plays a role in the confinement of quarks inside hadrons through black-hole analogies
\cite{Salam,Tennakone,Sivaram1,Sivaram2,Sivaram3}, although not within the framework 
of considering weak interactions as derivable from gravity at short lengthscale. 

We envisage here at least three experimental scenarios in which this ``Newton-Fermi conjecture,'' 
centered around Eq. (3), could give rise to novel observable effects. Specifically: 

\vspace{0.3cm}
\noindent
(a) {\sl High-precision measurements of the Newton gravitational constant in the micrometer 
to nanometer range} 

\vspace{0.1cm}
\noindent
Unless there is an abrupt dependence of $G_N$ on distance in the form of a resonant phenomenon, 
we expect that its increase with decreasing distance will be distributed over many distance decades. 
This means that evidences for $G_N$ as a running coupling constant could be achieved by comparing 
high-precision measurements at different distances, for instance in the cm, $\mu$m, and nm range. 
This provides further motivations to ongoing efforts to measure gravitationally-related forces 
in the submillimeter range \cite{Antoniadis}, especially as a byproduct of Casimir force studies, 
combining precision measurements and accurate theoretical predictions \cite{Onofrio}. 
It also provides motivations to improve the precision in the measurement of $G_N$ at macroscopic 
distances, adding on former discussions on its variability \cite{Long1,Long2,Long3}. 

\vspace{0.3cm}
\noindent
(b) {\sl Gravitationally bound states of hadrons and leptons in the subnanometer to femtometer range}

\vspace{0.1cm}
\noindent
Bound states may be sensitive to an anomalous gravitational coupling. The Rydberg constant is known 
from hydrogen spectroscopy with a precision of 5 parts per trillion, so this leads to bounds at  
the scale of $10^{-10}$ m on $G_N(10^{-10} \mathrm{m}) < 6 \times 10^{27} G_N < 2.4 \times 10^{-6} \tilde{G}_N$. 
With such a bound on $G_N$ in the nanometer range it seems very difficult to observe gravitationally 
bound dineutron states, for instance the Bohr radius should be of the order of 10 $\mu$m, with a 
ground state binding energy of 2.6 nK. By using reflecting surfaces as in \cite{Nesvi1}, we expect 
that neutrons of energy large enough to approach the surface within hundreds of femtometers should 
feel the gravitational attraction of the atoms on the surface, resulting in a different energy 
spectrum with respect to the one already measured with ultracold neutrons \cite{Nesvi2}. 
Other suitable states are provided by exotic atoms. The spectroscopy of antiproton-nucleus 
atoms  \cite{Hori} was used to give an upper bound $G_N(10^{-13} \mathrm{m}) < 1.3 \times 10^{28} G_N$ 
\cite{Nesvi3}.  Muonic hydrogen should be also sensitive because of the small Bohr radius of 
$2.5 \times 10^{-13}$ m. The recently observed anomalous Lamb shift, so far interpreted as a measurement 
of the proton radius in disagreement with the CODATA value \cite{Pohl}, could instead be due 
to different gravitational binding energy in the 2$s_{1/2}$ and 2$p_{1/2}$ states, requiring 
evaluation of the related contribution beyond perturbation theory.

\vspace{0.3cm}
\noindent
(c) {\sl Gravitationally bound states of heavy quarks}

\vspace{0.1cm}
\noindent
Gravitation should contribute to the binding energy of the nuclei more substantially 
if coupled through $\tilde{G}_N$. It is however easy to check that for the value of 
$\tilde{G}_N$ assumed above this contribution is still negligible even with respect 
to the electrostatic correction, and below the accuracy achieved so far in the comparison 
between theoretical semiempirical mass formulae and binding energies as determined via mass spectrometry. 
We instead expect that bound states of heavy quarks will get a significant contribution 
from anomalous gravitational energy. An estimate for the gravitational contribution to the binding energy 
of heavy quarkonia of mass $m_q$, proportional to $\tilde{G}_N m_q^2$, is naturally obtained through 
its comparison to the corresponding electrostatic contribution, proportional to 
$e^2/4\pi \epsilon_0$ via fractional charges $2/3$ and $1/3$ for top and bottom quarks, 
respectively. The situation seems favourable in the case of $t\bar{t}$ pairs, since the larger mass 
leads to gravitational binding energy {\sl larger} than the electromagnetic contribution by a factor 
of about 130 even taking into account the larger value of $\alpha_{\mathrm{em}}$ due to its running at the 
top mass scale. However, in this case the intrinsically short lifetime of the top quark 
may prevent an easy signature, also considering the limits in energy resolution intrinsic in 
the initial states using hadron colliders such as Tevatron and LHC. 
Nevertheless, it is intriguing to investigate if recent anomalies in the $t\bar{t}$ 
production at the Tevatron \cite{Aaltonen,Abazov} could find an explanation in 
this setting. Next to this, $b\bar{b}$ bound states should have instead 
an anomalous gravitational contribution about 3 times smaller than the electromagnetic one. 
In principle, high resolution spectroscopy of $b\bar{b}$ bound states could identify 
such a further contribution to the binding energy using future electron-positron 
collider machines optimized for bottom quarks such as SuperB and SuperKEKb, although 
extraction of this anomalous term requires an accurate mastering of the dominant binding 
energy due to color interactions. From this perspective, the best scenario is an 
electron-positron storage ring producing bound states of $\mu^+\mu^-$ \cite{Brodsky} 
or $\tau^+\tau^-$ \cite{Moffat,Avilez}, allowing one to study their spectroscopy at high precision.  

\vspace{0.1cm}

It may be worth to iterate that, as clearly emphasized in the conclusions of 
\cite{Capozziello1}, unification schemes based on attempts to geometrize weak 
interactions differ conceptually from the usual one pursued in the latest decades 
based on gauging internal symmetries. In the latter case the general idea is to incorporate 
the standard model into broader algebraic structures which, apart from the elegant, 
predictive, falsifiable example of SU(5) GUT, in general have more free parameters 
and make predictions predominantly at higher, still unexplored, energies. 
In this geometrization program instead, the idea is to strive for an {\sl economic} 
and {\sl falsifiable} description of natural phenomena - without introducing further 
degrees of freedom or free parameters - in which the Fermi constant loses its fundamental 
character and is considered as an effective coupling constant emerging from the 
short-distance geometry of quantum vacuum. 
We believe that such a geometrical approach to unification is closer in spirit to the successful 
unification pursued by Maxwell for electric and magnetic phenomena, and allows for 
tests in the whole range of energies explored so far - as discussed above - without 
providing ample margins for arbitrarily tuning free parameters to accommodate possible 
unsuccessful predictions at the currently explored energy scale.

In conclusion, we conjecture that weak interactions should be considered as empirical 
evidences of quantum gravity at the Fermi scale. The Fermi constant plays the role 
of an effective gravitational constant taking into account the presence of quantum 
fluctuations of the spacetime foam \cite{Wheeler,Hawking,Garay,Carlip}. 
The existence of violations of external symmetries such as parity, and internal symmetries 
such as flavor, or local nonconservation of the electric charge as witnessed by the presence of 
charged intermediate vector bosons like $W^{\pm}$, may be related in this framework to topology 
changes induced by quantum fluctuations through tunnelling among superselected spaces.
This could also provide a geometrical interpretation of the universality of weak interactions 
\`a la Cabibbo, {\it i.e.}, via CKM and PMNS unitary matrices for quarks and leptons, respectively.  
While at this stage we prefer to maintain an agnostic attitude towards the origin of the conjectured 
running of the Newtonian constant, we note that - besides extra-dimensions and moduli predicted 
by string theory - recently developed $f(R)$-gravity \cite{Capozziello} and Higgsless models 
\cite{Calmet} may provide natural settings for this effect.
A gravitational running coupling constant may be a concept less difficult to accept 
with respect to the existence of extra-dimensions since, unlike the latter for which we 
have collected null experimental evidence so far, we are certain about the reality of  
running coupling constants for all other fundamental interactions. 
Further implications and remaining challenging questions will be analysed in the 
future, in particular the role of neutral weak currents, the possible relationship 
between the spin-1 bosons $W^{\pm}$ and the spin-2 graviton, and more in general the 
consistent incorporation within this unified scenario of all the experimental facts 
successfully corroborating the electroweak model.

\end{document}